\documentclass[fleqn,usenatbib]{mnras}

\usepackage{newtxtext,newtxmath}

\usepackage[T1]{fontenc}

\DeclareRobustCommand{\VAN}[3]{#2}
\let\VANthebibliography\thebibliography
\def\thebibliography{\DeclareRobustCommand{\VAN}[3]{##3}\VANthebibliography}

\usepackage{graphicx}	
\usepackage{amsmath}	

\usepackage{multirow}
\usepackage{tablefootnote}
\usepackage{longtable}

\title[Discovery of double stars by Giovanni Battista Hodierna in 1654]{Discovery of double stars by Giovanni Battista Hodierna in 1654}

\author[J.\,González-Payo \& J.\,A.\,Caballero]{
J.\,González-Payo,$^{1,2}$\thanks{E-mail: fcojgonz@ucm.es}
and J.\,A.\,Caballero,$^{3}$
\\
$^{1}$Departamento de Física de la Tierra y Astrofísica \& IPARCOS-UCM (Instituto de Física de Partículas y del Cosmos de la UCM), Facultad de Ciencias Físicas,\\
Universidad Complutense de Madrid, 28040 Madrid, Spain\\
$^{2}$UNIE Universidad, Departamento de Ciencia y Tecnología, c/ Arapiles 14, 28015 Madrid, Spain\\
$^{3}$Centro de Astrobiología (CSIC-INTA), European Space Astronomy Centre, Camino Bajo del Castillo s/n, 28692 Villanueva de la Cañada, Madrid, Spain
}

\date{Accepted 2024 August 20. Received 2024 August 20; in original form 2024 July 12}

\pubyear{\the\year{}}

\begin{document}
\label{firstpage}
\pagerange{\pageref{firstpage}--\pageref{lastpage}}
\maketitle

\begin{abstract}
It has been generally accepted that the originators of the double star astronomy were Christian Mayer and William Herschel. We recovered the memory of the poorly known Italian astronomer Giovanni Battista Hodierna, who published the first catalogue of stellar binaries over a century before Mayer and Herschel. We analysed the fourth section of 1654 G.\,B. Hodierna's book ``\textit{De systemate orbis cometici deque admirandis coeli characteribus}''. There, Hodierna listed a dozen pairs of stars whose identification with modern star names had been lost for centuries. To identify the pairs, we used Hodierna's Latin descriptions of location in constellations for all primary stars, ecliptic coordinates and angular separations to companions for some, and the Washington Double Star, \textit{Hipparcos}, and \textit{Gaia} catalogues. We were able to identify the twelve primaries and eleven multiple systems with companions, of which nine were double and two were triple.   Besides, with up-to-date data, we confirmed that four systems are physically bound: Atlas and Pleione, $\alpha^{1,2}$~Lib, $\nu^{1,2}$~Dra, and $\theta^{1}$~Ori~A, C, and D. The other seven pairs are alignments of very bright stars at different distances.
\end{abstract}




\section{Introduction}
\label{sec:introduction}

For centuries, but specially during the last decades, stellar multiplicity has helped astronomers to learn about the formation, evolution, and parameters of stars \citep{duquennoy91,eggleton06,raghavan10,chabrier03,duchene13,tokovinin14}.
Although we now use powerful telescopes to investigate them, double stars have been observed since ancient times.
Ptolemy (c.100--c.170 AD) was the first astronomer to assign the term ``diplous'' ($\delta\iota\pi\lambda\textit{o}\upsilon\sigma$) to a double star, specifically $\nu^{1}$~Sgr and $\nu^{2}$~Sgr. 
These stars, separated by about 14\,arcmin, were observed with the naked eye, recorded in Ptolemy's Almagest, \textit{``H\=e Megal\=e (Math\=ematik\=e) Syntaxis''}\footnote{``The Great (Mathematical) Treatise''.}, and are now identified as an optical double (i.e., non physically bound). 
In old Arab astronomy, numerous star names collectively denoted two or more adjacent stars easily distinguishable to the naked eye -- actually, $\nu^{1}$~Sgr and $\nu^{2}$~Sgr were Ain al Rami, ``the eye of the Archer''.
Conversely, the widely recognised naked-eye duo Mizar ($\zeta$~UMa) and Alcor (g~UMa) within the Big Dipper garnered distinct names, likely due to their differing brightness levels, which did not portray them as a discernible ``pair'' of stars \citep{heintz78}.
Mizar and Alcor are not gravitationally bound either, although both of them are coeval members or the Ursa Major moving group \citep{riedel17,gagne18a}. 

It was not until the seventeenth century, with the invention of the telescope, that interest in double stars was rekindled. 
This led to the discovery of a new visual binary, Mizar A and B, attributed to the Italian mathematician Benedetto Castelli. 
Castelli had requested Galileo to observe it in 1616, although the discovery was later inaccurately credited to Giovanni Battista Riccioli in 1650 \citep{ondra04};
Mizar A and B form indeed a physical pair \citep{vogel1901}.
Galileo also resolved, on 4 February 1617, the three brightest stars of $\theta^{1}$~Ori, the central star of Orion's sword \citep{fedele49}. 
Besides, in his ``\textit{Almagestum Novum}'', \citet{riccioli1651} proposed another two pairs of stars separated by a few arcminutes in Capricorn and the Hyades (Sect.~\ref{sec:riccioli}). 

Previously to our work, the next registered reference to multiple stars in the history was from the Dutch scientist Christiaan Huygens, who published in 1659 the book \textit{``Systema Saturnium''} \citep{huygens1659}.
There, Huygens presented an engraving announcing that $\theta^{1}$~Ori was actually a group of three very close stars surrounded by an extended nebulous region. 
Galileo had also noted the trio 42 years earlier, although he had not recognised the Orion Nebula due to unknown reasons.
The trio of stars was lately extended to a quartet in 1673 by Jean Picard \citep{bond1848,wesley1900}.

Huygens opened a new way in the search of new double stars to other astronomers such as Robert Hooke, who detected {Mesarthim} ($\gamma$\,Ari) in 1664 \citep{argyle19}, or the two French Jesuit Fathers Jean de Fontaney and Jean Richaud, who discovered the duplicity of the stars {Acrux} ($\alpha$ Cru) in the Southern Cross, in 1685 from Cape Town (South Africa), and $\alpha$\,Cen, in 1689 from Pondicherry (India), respectively \citep{henroteau28,kameswararao84,kochhar91}. 
Other astronomers contributed to the knowledge of multiple stars, such as 
Giovanni Biachini, 
James Bradley, 
Jean Cassini, 
Gottfried Kirch, 
Nevil Maskelyne, 
Charles Messier, 
John Michell, 
and Nathaniel Pigott, 
until the arrival of the first catalogue of double stars.
Published in 1779 by Christian Mayer, court astronomer at Mannheim, his book \textit{``De novis in coelo sidereo phaenomenis in miris stellarum fixarum comitibus''}\footnote{``On new phenomena in the starry sky among the amazing companions of the fixed stars.''} contained in its last pages the Tabula Nova Stellarum Duplicium, that is, the new table of double stars \citep{schlimmer07}. 
The first version of the catalogue by \citet{mayer1779} contained 72 double stars, which increased to 80 with the new version by \citet{bode1781} two years later.
Mayer's Tabula Nova Stellarum Duplicium is considered to be the first catalogue of double stars.

The objects within a double system were not initially considered as connected as the prevailing belief was that it was merely a coincidental occurrence. 
According to \citet{schlimmer07}, William Herschel heard about Mayer’s work and started his own observations of double stars.
Herschel published three catalogues of possible double stars: ``The Catalogue of Double Stars'' \citep{herschel1782} with 269 entries, ``Catalogue of Double Stars'' \citep{herschel1785} with 484 entries, and ``On the Places of 145 New Double Stars'' \citep{herschel1822}. 
When Herschel reexamined his first catalogue in 1802, after claims of common proper motion and orbital motion by \citet{mayer1786}, he confirmed that several double stars were gravitationally bound, which sparked modern astronomy \citep{williams14}.

In his PhD thesis, \citet{longhitano11} noticed that, between the discoveries of Castelli and Galileo in 1616--1617 and the publication of the book by Huygens in 1659, and over a century before the double star catalogues of Mayer and Herschel, a poorly known Sicilian astronomer, Giovanni Battista Hodierna, published a short list of possible binary systems that did not leave much of a mark on history and have gone since unnoticed. 
We focus on the seminal work of Hodierna by uncovering documentation of some pairs of stars deemed binary, a revelation potentially predating many celebrated observations. 
These celestial twins, concealed within Hodierna's principal opus, offer a compelling narrative of early astronomical inquiry. 
Our investigation aims to shed light on these overlooked observations, highlighting their significance in the annals of scientific discovery, and underscoring their potential to reshape our understanding of early modern astronomy.

\section{A historical review}
\label{sec:historical_review}

\subsection{Giovanni Battista Hodierna}
\label{sec:GB_Hodierna}

Giovanni Battista Hodierna was born on 13 April 1597 in Ragusa, Italy. 
Little is known about his early life and education, but that he studied at the University of Palermo, where he likely developed his interest in astronomy and mathematics. 
He became a catholic priest and served as such in the then Kingdom of Sicily. 
During his religious life, Hodierna dedicated much of his time to observing the night sky. His main focus was on cataloguing stars and creating celestial maps. 
Hodierna made his observations using rudimentary instruments, such as refracting telescopes of his own design. 
Despite the limitations of his tools, he managed to identify and catalogue numerous stars and celestial objects, some of which had not been documented before. 
Hodierna is particularly known for his star catalogue, entitled ``\textit{De systemate orbis cometici deque admirandis coeli caracteribus}\footnote{``On the system of the cometary universe and on the admirable characteristics of the sky.''}'' \citep{hodierna1654}. 
The work anticipated Messier's work, but had little impact, and neither Messier nor any European astronomer seem to have known of it \citep{serio85}. 
Therefore, Hodierna never reached the same fame as other contemporary astronomers, such as Galileo Galilei or Johannes Kepler. 
After his death on 6 April 1660 in Palma di Montechiaro, his work was largely forgotten and rediscovered only in later centuries when modern astronomers recognised the importance of his observations \citep{serio85,jones86}.

\subsection{``\textit{De systemate orbis cometici}''}
\label{sec:systemate}


For our investigation, we used the ETH-Bibliothek Z\"urich copy of ``\textit{De systemate orbis cometici}'' (Rar 2876).
The book is structured in two parts divided in their turn into four sections. 
The first part refers to the theory of comets. Following the path outlined by Galileo, Hodierna distinguished the nature of comets from that of nebulae, attributing an earthly nature to the former and recognising only a celestial, stellar nature to the latter \citep{pavone86}. 
In the second part, ``\textit{De admirandis coeli caracteribvs}''\footnote{``On the admirable characteristics of the sky''.}, he classified and catalogued deep sky objects, which he denominated nebulae \citep{jones91}.
There, he tabulated and provided finding charts for a number of prominent objects, including the $\alpha$~Persei open cluster (M20), the Butterfly Cluster (M6), or the Lagoon Nebula (M8), over a century before their independent discovery by Jean-Philippe Loys de Cheseaux, Guillaume Le Gentil, Charles Messier, or Caroline Herschel, just to mention some names \citep{serio85,williams14}.
The fourth section of the second part, entitled ``\textit{In qua de stellis contiguis duplicibus seu Geminis deque Mundani Systematys Coperniceorum implicantia ratiocinandum venit}''\footnote{``Wherein the discourse concerning close double stars, or twins, and the implications for the Copernican system of the Universe, is to be considered.''} includes the list of binary stars that we investigated here, and for which he used the word ``geminae'' (twins).

\subsection{``Stellae geminae''}
\label{sec:description_binaries_text}

\begin{table*}
 \centering
 \caption[]{Hodierna's star pairs with coordinates.}
 \begin{tabular}{c@{\hspace{2mm}}l@{\hspace{2mm}}l@{\hspace{3mm}}l@{\hspace{2mm}}c@{\hspace{2mm}}c@{\hspace{4mm}}c@{\hspace{2mm}}c@{\hspace{2mm}}c@{\hspace{2mm}}l@{\hspace{2mm}}c@{\hspace{2mm}}c}
 \hline
 \noalign{\smallskip}
Id. & Star name  & Translation &  \multicolumn{3}{c}{Ecliptic longitude} & \multicolumn{3}{c}{Ecliptic latitude} & $\rho_{\rm text}$ & Segment & $\rho_{\rm drawing}$ \\
  &  & to English & Signum  & (deg) & (min) & (deg) & (min) & Hemis. & (arcmin) &  & (arcmin) \\
 \noalign{\smallskip}
 \hline
 \noalign{\smallskip}
\multirow{2}{*}{1} & \multirow{2}{*}{Orientalissima Pleiadum} & The easternmost & \multirow{2}{*}{Tauri} & \multirow{2}{*}{25} & \multirow{2}{*}{32} & \multirow{2}{*}{3} & \multirow{2}{*}{52} & \multirow{2}{*}{B} & \multirow{2}{*}{Three} & \multirow{2}{*}{EF} & \multirow{2}{*}{3.8} \\
 & &  of the Pleiades  & & & & & & & & & \\
 \noalign{\smallskip}
\multirow{2}{*}{2} & \multirow{2}{*}{Oculus Boreus Tauri} & The northern eye & \multirow{2}{*}{Gemin.} & \multirow{2}{*}{3} & \multirow{2}{*}{39} & \multirow{2}{*}{2} & \multirow{2}{*}{36} & \multirow{2}{*}{A} & \multirow{2}{*}{Almost five} & \multirow{2}{*}{GH} & \multirow{2}{*}{4.6} \\
 & & of the bull & & & & & & & & & \\
 \noalign{\smallskip}
\multirow{2}{*}{3} & \multirow{2}{*}{Lanx Austrina Libr\ae} & Southern pan & \multirow{2}{*}{Scorpionis} & \multirow{2}{*}{10} & \multirow{2}{*}{14} & \multirow{2}{*}{0} & \multirow{2}{*}{26} & \multirow{2}{*}{B} & \multirow{2}{*}{Three} & \multirow{2}{*}{CD} & \multirow{2}{*}{3.0} \\
 & & of the balance  & & & & & & & & & \\
 \noalign{\smallskip}
\multirow{2}{*}{4} & Cornu Occidentale Capric. & The western horn & \multirow{2}{*}{Capric.} & \multirow{2}{*}{29} & \multirow{2}{*}{4} & \multirow{2}{*}{7} & \multirow{2}{*}{2} & \multirow{2}{*}{B} & Almost five & \multirow{2}{*}{IK} & \multirow{2}{*}{5.2} \\
 & (In praecedenti Cornu Capricorni) &  of the capricorn & & & & & & & and a half & & \\
 \noalign{\smallskip}
\multirow{2}{*}{5} & \multirow{2}{*}{Trium in frontem Occid.} & Three towards & \multirow{2}{*}{Scorpionis} & \multirow{2}{*}{28} & \multirow{2}{*}{20} & \multirow{2}{*}{1} & \multirow{2}{*}{40} & \multirow{2}{*}{B} & \multirow{2}{*}{...} & \multirow{2}{*}{...} & \multirow{2}{*}{...} \\
 & & the western front & & & & & & & & & \\
 \noalign{\smallskip}
 \hline
 \end{tabular}
\label{tab:hodierna_table_with_coordinates}
\end{table*}

\begin{figure}
  \centering
  \includegraphics[width=1\linewidth, angle=0]{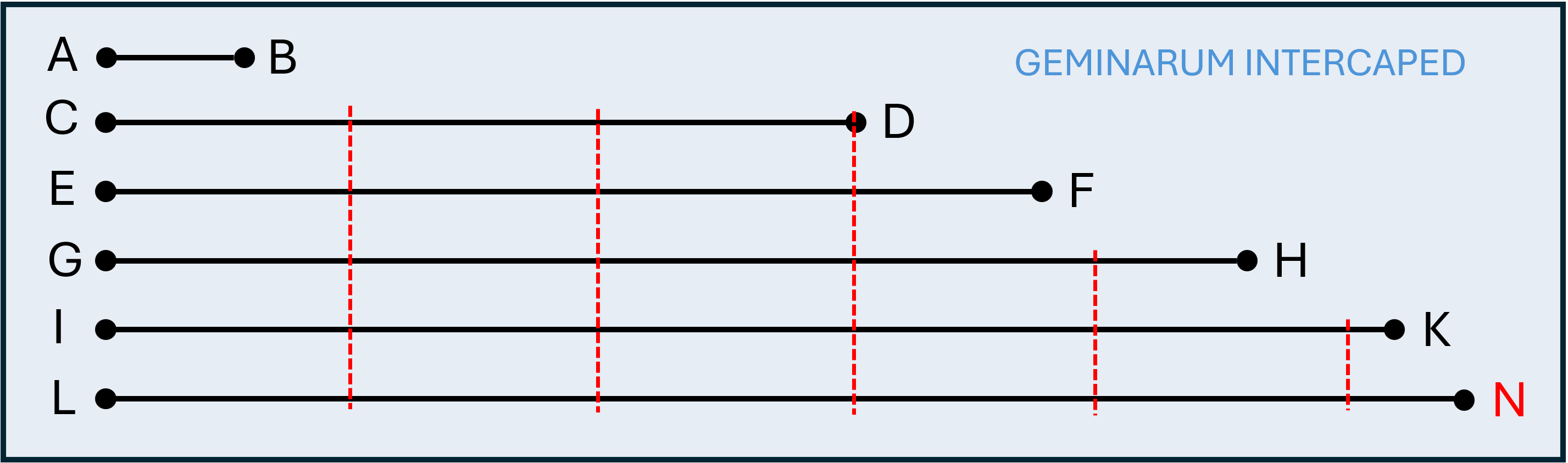}
 \caption[]{Adaptation of Hodierna's separations drawings. 
 Each segment corresponds to an angular separation.
 The vertical red lines are 1\,arcmin ticks. 
 The red N denotes a typographical error (the segment LN is referred in Hodierna's book as LM).}
  \label{fig:geminarum}
\end{figure}

At the beginning of the section ``\textit{In qua de stellis contiguis duplicibus}'', Hodierna stated some sentences that were very advanced for their epoch: 
``everywhere throughout the ether, certain binary stars shine, which, although double in themselves, are so closely bound by a bond of contiguity that hardly two are distinguished from each other, by the bond of contiguity with which they appear to be connected, and thus are not considered double, but wholly single [...] but many, or countless, such double stars are found throughout the vastness of the sphere, as scarcely a constellation in the sky shines in which there is not at least one or two binaries, especially in the luminous nebulous regions and in the dark stretches of the sky, among which there are some very prominent ones that adhere to the ecliptic in the Zodiac''. 
Following, the author included a table containing five binary stars that he called ``\textit{Stellae geminae iuxta eclypticam}''\footnote{``Binary stars near the ecliptic''.}.
We show an adaptation of Hodierna's table in the Table~\ref{tab:hodierna_table_with_coordinates}.
For each of the five stars, he provided a Latin name, which is often a very brief description of its location within its constellation (for example, \textit{``Oculus Boreus Tauri''} is ``the northern eye of the bull''), and the ecliptic coordinates in degrees and minutes of latitude and longitude (see below).
The longitude includes the ``signum'', which is the zodiacal area of the sky (not necessarily coinciding with the current constellation definition), and the latitude includes an additional character for the ecliptic hemisphere, namely ``B'' for boreal (north) and ``A'' for austral (south).
We gave the identification numbers \#1 to \#5 to the five stars.

After the table, Hodierna enumerated in the text seven binary stars for which he did not initially provide any quantitative datum, but only their Latin names.
We gave the Ids. \#6--\#12 to the seven stars.
Next, Hodierna displayed a drawing labelled ``\textit{Geminarum intercaped}'' (``Binaries separation'') with a pictographical representation of the angular separation between components of another seven binary systems.
We display in Fig.~\ref{fig:geminarum} our own adaptation of Hodierna's drawing. 
The ticks on the six segments indicate arcminutes.
There are six segments in Fig.~\ref{fig:geminarum} because the shortest one, of less than one arcminute, corresponds to two close binary stars that had not been mentioned before.
The seven binary systems for which Hodierna was able to measure separations with his rudimentary telescope were enumerated below the drawing, together with the approximate value of the angular separation in Latin. 
They have an entry in the last three columns of Table~\ref{tab:hodierna_table_without_coordinates}.
We provide both the translation of the angular separation in the text, $\rho_{\rm text}$, and our interpretation of the angular separation in the drawing, $\rho_{\rm drawing}$, with an uncertainty of 0.2\,arcmin, as they did not coincide exactly.
At least four of the seven stars were already in Hodierna's table, and the names of the other three had been enumerated in the text.
Finally, although Hodierna did not identify any more multiple systems, he emphasised the multitude of binary stars that were left out of his description: ``[...] and many others, whose catalogue would grow to infinity if they were individually noted''.

\begin{table*}
 \centering
 \caption[]{Hodierna's star pairs without coordinates.}
 \begin{tabular}{c@{\hspace{2mm}}l@{\hspace{4mm}}l@{\hspace{4mm}}lcc}
 \hline
 \noalign{\smallskip}
Id. & Star name &  Translation to English & $\rho_{\rm text}$ & Segment & $\rho_{\rm drawing}$ \\
  &  &   & (arcmin) &  & (arcmin) \\
 \noalign{\smallskip}
 \hline
 \noalign{\smallskip}
 6 & In pede sinistro pr\ae cedentis Geminorum & On the left foot of the preceding twin & ... & ... & ... \\
 \noalign{\smallskip}  
 7 & Ceruice Leonis  & At the tail of the lion & ... & ... & ...  \\
  \noalign{\smallskip}
 8 & In ancone al\ae\, dextr\ae\, Cygni  & On the right wing of the swan & ... & ... & ... \\
  \noalign{\smallskip}  
 9 & In pede posteriori sinistro Leporis & On the left hind leg of the hare & ... & ... & ... \\
  \noalign{\smallskip}
 10 & Secunda spondili Scorpionis & In the second vertebra of the scorpion & Five and a half & LM  & 5.4 \\ 
  \noalign{\smallskip}  
 \multirow{2}{*}{11} & Capitis Draconis, quatuor rombum &  On the head of the dragon, of the four that constitute  & \multirow{2}{*}{Less than one} & \multirow{2}{*}{AB} & \multirow{2}{*}{0.6} \\
   & constituentium, qu\ae\, sub Oculo exigua & the diamond, the one that is slightly below the eye & & & \\
  \noalign{\smallskip}  
 12 & Medie ensis Orionis & In the middle of Orion's sword & Less than one & AB & 0.6 \\ 
 \noalign{\smallskip} 
 \hline
 \end{tabular}
\label{tab:hodierna_table_without_coordinates}
\end{table*}

\section{Analysis and results}
\label{sec:hodierna_text_analysis}

We proceeded to the identification of the dozen stars mentioned by Hodierna and their companions using the coordinates, properties, and clues provided in his book.
First, for the five stars in Hodierna's table for which he provided ecliptic coordinates (Table~\ref{tab:hodierna_table_with_coordinates}), we computed their current equatorial coordinates.
For that, we first translated Hodierna's coordinates to a modern format using the relations of \citet{verbunt11}:

\begin{equation}
    \lambda=(Z-1)\cdot 30 + G_{\text{lon}} + \frac{M_{\text{lon}}}{60}
\end{equation}

\noindent and

\begin{equation}
    \beta=\pm \,\, G_{\text{lat}} + \frac{M_{\text{lat}}}{60},
\end{equation}

\noindent where $\lambda$ and $\beta$ are the ecliptic coordinates, $G_{\text{lon}}$ and $M_{\text{lon}}$ the degrees and minutes of longitude, $G_{\text{lat}}$ and $M_{\text{lat}}$ the degrees and minutes of latitude, $Z$ is the number of order for the zodiacal ``signum'' (starting with 1 for Aries and finishing with 12 for Pisces), and the sign of $\beta$ is positive for northern longitudes (borealis) or negative for southern longitudes (australis).
Second, we transformed ecliptic to equatorial coordinates with the NASA/IPAC Extragalactic Database Coordinate Calculator tool\footnote{\url{https://ned.ipac.caltech.edu/coordinate_calculator}}.
Third, we searched for the closest naked-eye stars that fit to the Latin name description.
Fourth, we measured the angular separation between the computed and current J2000 coordinates.
The mean absolute separation of the five identified stars is 4.838$\pm$0.063\,deg, which is identical to the expected offset by the equinox precession since 1654 of 4.837\,deg. 
For the computation of the equinox precession we used the equation of \citet{capitaine03}:
\begin{equation}
    p_A = 5\,028.796195\cdot T + 1.1054348 \cdot T^2 
\end{equation}
\noindent where $T$ is time in Julian centuries and $p_A$ is the absolute value of equinox precession in arcsec. 
The offset vectors in right ascension and declination were also very similar in the five cases.
Given the uncertainty in Hodierna's input coordinates, of arcminutes, and the relatively low proper motion of the stars, we did not take into account any proper motion correction.
The five identified stars are so bright that they have proper names: Atlas (\#1), Ain (\#2), Zubenelgenubi (\#3), Algedi (\#4), and Acrab (\#5). 
Furthermore, the location of the five stars correspond to their descriptive Latin names 
(e.g., Atlas is ``the easternmost of the Pleaides'').
They are listed in the top part of Table~\ref{tab:identified_pairs}.

\begin{figure}
  \centering
  \includegraphics[width=1\linewidth, angle=0]{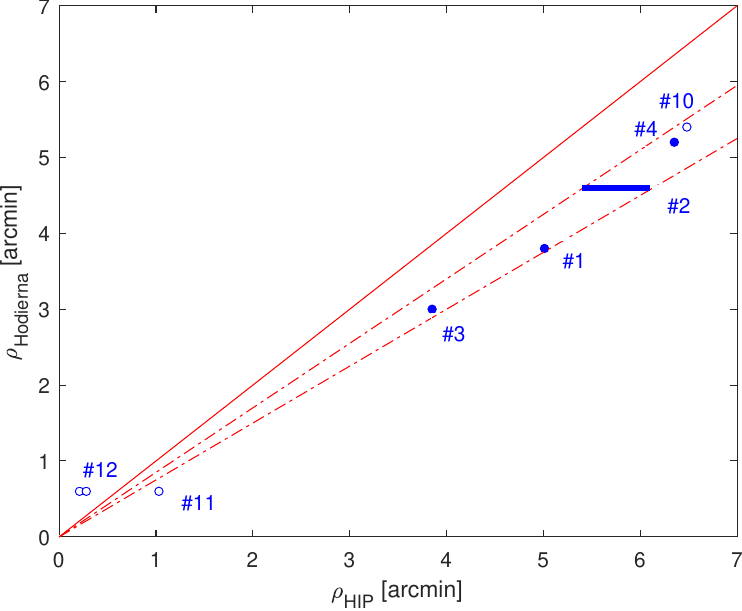}
 \caption[]{Binary angular separations $\rho$ supplied by Hodierna (when available) vs. $\rho$ calculated by us from \textit{Hipparcos} data. 
 Filled and empty blue dots are for stars with and without coordinates, respectively. The red solid line has a slope equal to 1, while the red dashed lines have slopes of 0.75 and 0.85. 
 The horizontal solid blue line represents the possible real values of separation that the companion of Ain should have in case that it existed.} 
  \label{fig:rho}
\end{figure}

Next, we used the Aladin sky atlas \citep{bonnarel00} and the Washington Double Star catalogue \citep[WDS;][]{mason01} to search for bright stellar companions to the five stars.
Since Hodierna's telescope was rudimentary, such companions should have a magnitude comparable to those of their primaries (a modern telescope with an aperture of 35\,mm, such as one for children, can detect stars up to  9.9\,mag with a 5\,mm exit pupil; \citealt{north14}).
Four of the five stars had two angular separations measured by Hodierna, namely one from the text and the other from the segment drawing; hereafter, we used only the angular separation from the drawing, dubbed $\rho_{\rm drawing}$ (Fig.~\ref{fig:geminarum} and last column of Table~\ref{tab:hodierna_table_with_coordinates}).
We were able to identify naked-eye companions to three of the four stars.
The three pairs are Atlas and Pleione (\#1, the easternmost of the Pleiades),
Zubenelgenubi and $\alpha^{1}$ Lib (\#3, the southern weighing pan of Libra), 
and Algedi and Prima Giedi\footnote{See Sect.~\ref{sec:riccioli} for a potential previous discovery of the Algedi and Prima Giedi pair by Martinus Hortensius as described in Riccioli's \textit{``Almagestum Novum''}.
Prima Giedi should not be confused with Giedi Prime.} (\#4, the western horn of the goat). 
Fig.~\ref{fig:rho} shows a comparison of Hodierna's $\rho_{\rm drawing}$ and our own measurements using \textit{Hipparcos} data \citep{perryman97} computed with spherical astronomy equations as \citet{gonzalezpayo23}.
From the comparison of the three pairs, Hodierna measured smaller angular separations by a $\sim 0.80$ scale factor.
The $V$-band magnitudes of the primaries and secondaries range in the 2.8--3.6\,mag and 4.3--5.2\,mag intervals, well within the naked eye limit with typical dark sky conditions.  

We were not able to identify any suitable companion to the fourth star with ecliptic coordinates and $\rho_{\rm drawing}$, namely Ain (\#2, $\epsilon$~Tau).
From Fig.~\ref{fig:rho}, we expected such a suitable companion at an actual angular separation of about 5.4--6.1\,arcmin, from Hodierna's value of 4.6\,arcmin and $\sim$ 0.80 scale factor.
However, there are no stars brighter than 13.5\,mag in the wider 5--7\,arcmin annulus centred on Ain. 
Trying to ascertain whether Hodierna's claim of a companion to the bright star in Taurus, not far from the ecliptic ($\beta \approx -2.6$\,deg), was the result of an observational mistake (e.g. an optical ghost) or an actual measurement (e.g. a bright Solar System body passing near the ecliptic) would be just speculating.
Nevertheless, we were confident of our identification, because Ain is indeed ``the northern eye of the bull'' in many classical sky descriptions (e.g. ``\textit{Reliqua quae est in oculo boreali}'' in Ptolemy's Almagest, ``\textit{Oculus boreus}'' in \citealt{flamsteed1725}), and the offset between precession-corrected Hodierna's coordinates and ours in J2000 is only slightly less of 1\,arcmin.

The fifth star in Table~\ref{tab:hodierna_table_with_coordinates}, Acrab (\#5), does not have a Hodierna's $\rho_{\rm drawing}$ but is the only naked eye star in the trio of the western front (claws) of the Scorpion that has a relatively bright companion at an angular separation comparable to those of the pairs \#1, \#3, and \#4. 
The most probable companion is HD~144273, which has a visual magnitude of 7.54\,mag and is separated to Acrab by 8.65\,arcmin.

Next, we went on identifying the rest of stars without ecliptic coordinates.
First, we searched for bright stars that match the Latin descriptions of pairs \#10, \#11, and \#12 (see the translation to English in Table~\ref{tab:hodierna_table_without_coordinates}) and have relatively bright companions at angular separations similar to those estimated by Hodierna.
The two closest pairs, with $\rho_{\rm drawing}$ of ``less than one [arcminute]'' according to Hodierna, are Kuma$^{2}$ ($\nu^{2}$ Dra) and Kuma$^{1}$ ($\nu^{1}$ Dra) (\#11), whose components are actually separated by 1.03\,armin, and $\theta^{1}$ Ori (\#12).
The later is the famous asterism of the Trapezium \citep{herbig86,mccaughrean94,simondiaz06}.
Hodierna may be able to see the three brighest components of the Trapezium, namely $\theta^{1}$ Ori A, C, and D, which had been already resolved by Galileo in 1617.
The separation between the eastern- and westernmost OB-type components of the Trapezium is about 0.2\,arcmin, probably close to the resolution limit of Hodierna's instrumentation and site.
The last pair with $\rho_{\rm drawing}$ is very likely $\zeta^{2}$~Sco and $\zeta^{1}$~Sco (\#10), which is made of two stars brighter than $V$ = 5\,mag separated by 6.48\,arcmin.
Hodierna's measurement of their separation was also affected by the same scale factor $\sim0.80$ as for the pairs \#1, \#3, and \#4 (Fig.~\ref{fig:rho}), which strengthen our identification.

Finally, also guided by the Latin description and existence of relatively bright companions at a few arcminutes to naked-eye stars, we tentatively identified the four remaining pairs (\#6, \#7, \#8, and \#9).
Save for one case, the components are separated by 3.9--6.5\,arcmin.
The exception is the pair $\nu$ Gem and HD~257937 (\#6), which are separated by only 1.9\,arcmin.
Similarly, the companions are probably at the faintest limit of Hodierna's instrumentation, at $V \approx$ 8.2--9.0\,mag, save again for one exception: the $\theta$ Cyg system (\#8).
This system has in common with the Orion's Trapezium that Hodierna may have seen three components, since R~Cyg and HD~185264 are more of less of the same visual magnitude, $V \approx$ 6.1--6.5\,mag, and at the same angular separation to $\theta$~Cyg.

\begin{table*}
 \centering
 \caption[]{All star pairs described by Hodierna.}
 \begin{tabular}{c@{\hspace{3mm}}l@{\hspace{3mm}}c@{\hspace{3mm}}c@{\hspace{3mm}}c@{\hspace{3mm}}c@{\hspace{3mm}}c@{\hspace{3mm}}c@{\hspace{3mm}}c@{\hspace{3mm}}c@{\hspace{3mm}}c}
 \hline
 \noalign{\smallskip}
Id & Star name & $\alpha$ (J2000) & $\delta$ (J2000) & $\mu_{\alpha}\cos\delta$ & $\mu_\delta$ & $d$ & $V$ & WDS & Disc. code & $\rho$ \\
  &  &  (hh:mm:ss.ss) & (dd:mm:ss.s) & (mas\,a$^{-1}$) & (mas\,a$^{-1}$) & (pc) & (mag) &   &  & (arcmin) \\
 \noalign{\smallskip}
 \hline
 \noalign{\smallskip}
 1 & Atlas (27 Tau) & 03:49:09.74 & +24:03:12.3 & 21.761 & --46.241 & 118.6 & 3.63 & \multirow{2}{*}{...} & ... & \\
   & Pleione (28 Tau) & 03:49:11.22 & +24:08:12.2 & 19.496 & --47.650 & 138.1 & 5.09 &  & ... & 5.01 \\  
 \noalign{\smallskip}
 \hline
 \noalign{\smallskip}  
 2 & Ain ($\epsilon$ Tau) & 04:28:37.00 & +19:10:49.6 & 107.53 & --36.200 & 44.71 & 3.53 & \multirow{2}{*}{...} & ... & \\  
   & Not identified & ... & ... & ... & ... & ... & ... &  & ... & ... \\  
 \noalign{\smallskip}
 \hline
 \noalign{\smallskip}  
 3 & Zubenelgenubi ($\alpha^{2}$ Lib) & 14:50:52.71 & --16:02:30.4 & --105.68 & --68.400 & 23.24 & 2.75 & \multirow{2}{*}{14509--1603} & SHJ 186 A &  \\
   & $\alpha^{1}$ Lib & 14:50:41.17 & --15:59:50.0 & --73.505 & --67.871 & 23.94 & 5.16 &   & SHJ 186 B & 3.85 \\  
 \noalign{\smallskip}
 \hline
 \noalign{\smallskip}   
 4 & Algedi ($\alpha^{2}$ Cap)  & 20:18:03.26 & --12:32:41.5  & 61.212 & 2.4120 & 33.43 & 3.58 & \multirow{2}{*}{20181--1233} & STFA 51 A &  \\
   & Prima Giedi ($\alpha^{1}$ Cap) & 20:17:38.87 & --12:30:29.6 & 21.709 & 1.6430 & 249.0 & 4.27 &  & STFA 51 E & 6.35 \\  
 \noalign{\smallskip}
 \hline
 \noalign{\smallskip}  
 5 & Acrab ($\beta$ Sco) & 16:05:26.23 & --19:48:19.4 &  --5.200 & --24.040  & 123.9 & 2.50 & \multirow{2}{*}{...} & ... &  \\  
  & HD 144273 &  16:05:44.84 & --19:40:51.8  & --10.160 & --21.824  & 152.7 & 7.54 &  & ... & 8.65 \\
 \noalign{\smallskip}
 \hline
 \noalign{\smallskip}  
6 & $\nu$ Gem & 06:28:57.79 & +20:12:43.7 & --4.669 & --14.402	& 224.8 & 4.14 & \multirow{2}{*}{06290+2013} & STTA 77 A & \\
   & HD 257937 & 06:28:53.77 & +20:14:21.2 & --3.121 & --7.624	& 122.8 & 8.97 &  & STTA 77 B & 1.88 \\  
 \noalign{\smallskip}
 \hline
 \noalign{\smallskip}  
7 & Denebola ($\beta$ Leo) & 11:49:03.58 & +14:34:19.4 & --497.68 & --114.67 & 11.00 & 2.13 & \multirow{2}{*}{11491+1434} & BU 604 A &  \\
 & BD+15 2382 & 11:48:59.18 & +14:30:27.2 & --45.481 & 31.050 & 117.9 & 8.46 &  & BU 604 D & 4.01 \\  
 \noalign{\smallskip}
 \hline
 \noalign{\smallskip}  
 8 & $\theta$ Cyg & 19:36:26.53 & +50:13:16.0 & --5.854 & 263.66 & 18.43 & 4.48 & \multirow{3}{*}{...} & ... &  \\
  & HD 185264$^{a}$ & 19:35:55.95 & +50:14:19.0 & --2.805 & 36.659 & 161.3 & 6.46 &  & ... & 4.99 \\
  & R Cyg$^{a}$ & 19:36:49.36 & +50:11:59.7 & --4.194 & --6.2730	& 597.1 & 6.10 &  & ... & 3.89 \\
 \noalign{\smallskip}
 \hline
 \noalign{\smallskip}   
 9 & $\delta$ Lep & 05:51:19.30 & --20:52:44.7 & 229.03 & --647.93 & 35.18 & 3.85 & \multirow{2}{*}{...} & ... &  \\  
   & HD 39405 & 05:51:36.77 & --20:50:13.0 & --4.186 & --21.897 & 315.5 & 8.25 &  & ... & 4.84 \\ 
 \noalign{\smallskip}
 \hline
 \noalign{\smallskip}  
  10 & $\zeta^{2}$ Sco & 16:54:35.01 & --42:21:40.7 &  --126.72 & --228.84 & 41.26 & 3.62 & \multirow{2}{*}{...} & ... &  \\  
  & $\zeta^{1}$ Sco & 16:53:59.73 & --42:21:43.3 & --0.0940 & --3.3680 & 1\,708 & 4.79 &  & ... & 6.48 \\   
 \noalign{\smallskip}
 \hline
 \noalign{\smallskip}   
 11 & Kuma$^{2}$ ($\nu^{2}$ Dra) & 17:32:16.04 & +55:10:22.5 & 144.34 & 59.585 & 29.92 & 4.87 & \multirow{2}{*}{17322+5511} & STFA 35 A & \\
   & Kuma$^{1}$ ($\nu^{1}$ Dra) & 17:32:10.57 & +55:11:03.3 & 148.04 & 54.194 & 30.11 & 4.90 &  & STFA 35 B & 1.03 \\  
 \noalign{\smallskip}
 \hline
 \noalign{\smallskip}  
 12 & $\theta^{1}$ Ori C & 05:35:16.50 & --05:23:22.9 & 2.262 & 0.994 & 399.8 & 5.13 & \multirow{3}{*}{05353--0523} & STF 748 C & \\  
   & $\theta^{1}$ Ori A$^{b}$ & 05:35:15.83 & --05:23:14.3 & 1.355 & 0.250 & 378.4 & 6.73 &  & STF 748 A & 0.21 \\  
   & $\theta^{1}$ Ori D$^{b}$ & 05:35:17.26 & --05:23:16.6 & 1.822 & 0.393 & 438.21 & 6.70 &   & STF 748 D & 0.22 \\ 
 \noalign{\smallskip} 
 \hline
 \end{tabular}
 \label{tab:identified_pairs}
 \begin{flushleft}
    \textit{Notes.} $^{a}$ Both stars are optical companions of $\theta$ Cyg.
    $^{b}$ Both stars are physical companions of $\theta^{1}$ Ori C. The three stars A, C, and D make up the Orion's Trapezium.
 \end{flushleft}  
\end{table*}

Table~\ref{tab:identified_pairs} collects the results and main properties of the 12 systems, including equatorial coordinates, proper motions, distances from \textit{Hipparcos} and \textit{Gaia}, and visual magnitudes from SIMBAD.
There are actually two triples and nine binaries, for which we provide our angular separations in the last column, and one single (Ain).
The table also shows the WDS identifiers and discovery codes of six systems, assigned to F.\,G.\,W.~Struve (STF, STFA), J.~South \& J.~Herschel (SHJ), S.\,W.~Burnham (BU), and O.~Struve (STTA) instead of to Hodierna (or even to Galileo and Castelli in the case of the Orion's Trapezium, or to Hortensius and Riccioli for Algedi and Prima Giedi; Sect.~\ref{sec:riccioli}).
The 12 primary stars and their companions are also illustrated by the thumbnails in Fig.~\ref{fig:binary_stars_images}.

Not all of the eleven identified systems are physically bound.
Furthermore, there are optical systems with components at very different heliocentric distances.
For example, the system \#8 is made of an F3 dwarf at 18.4\,pc ($\theta$ Cyg), a G9\,III eruptive variable at $\sim$160\,pc (HD~185264), and a S-type star at $\sim$600\,pc (R~Cyg).
We applied the conditions outlined by \citet{gonzalezpayo23} for common proper motion and parallax systems, but accounting for the greater uncertainties in parallactic distances of the brightest stars.
Since they are too bright for \textit{Gaia}, we took their parallax values from \textit{Hipparcos}.
After that, there remain four systems that are actually bound or are part of the same astrophysical association.
Two of them are in young open clusters, namely Atlas and Pleione (\#1) in the Pleaides and $\theta^{1}$~Ori (\#12) in the Orion Nebula Cluster.
A third bound system, $\alpha$~Lib (\#3) is also young and has been proposed to belong to the contested $\sim$300\,Ma-old Castor stellar kinematic group \citep{barrado98,montes01,mamajek13,zuckerman13} 
Regardless of the actual membership of $\alpha$~Lib in the group, the stellar system is indeed young based on a dusty disc and, possibly, mid-infrared emission features around the primary \citep{chen05}.
The system is actually quadruple, since both $\alpha^{2}$~Lib and $\alpha^{1}$~Lib are double themselves ($\alpha^{2}$~Lib: \citealt{slipher1904,lee1914,young1917,wilson53} -- $\alpha^{1}$~Lib: \citealt{duquennoy91,beuzit04,makarov05}).
Furthermore, \citet{caballero10} proposed a fifth component in the system at an extremely wide separation, KU~Lib, which also displays youth features such as high lithium abundance, strong X-ray emission, fast rotation, and photospheric dark spots \citep{gaidos98,gaidos00}.
The fourth and last bound system, namely Kuma$^{2}$ and Kuma$^{1}$ (\#11), is also triple (Kuma$^{2}$ is a spectroscopic binary -- \citealt{batten78,pourbaix04}), but we did not find any reliable reference in the literature for their membership in any stellar kinematic group.
Accounting for the reported masses of all the stellar components, all four systems have binding energies greater than 10$^{35}$\,J and up to 10$^{38}$\,J, between three and six orders of magnitude greater than those of the most fragile multiple systems \citep{caballero09,gonzalezpayo23}.  
The other seven identified pairs (namely Algedi [\#4], Acrab [\#5], $\nu$~Gem [\#6], Denebola [\#7], $\theta$~Cyg [\#8], $\delta$~Lep [\#9], and $\zeta^{2}$~Sco [\#10]) are optical and, therefore, are not bound.

\section{Conclusions}
\label{sec:conclusions}

It is established that Christian Mayer published the first catalogue of possible double stars in 1779, just a few years ahead of the three catalogues of William Herschel. 
Hence, both Mayer and Herschel inaugurated the double star astronomy.
However, here we demonstrate that the Italian astronomer Giovanni Battista Hodierna identified 12 double (and triple) stars in his ``\textit{De systemate orbis cometici}'', which dates back to 1654.
Hodierna measured angular separations for a few of them and remarked that there were many more multiple systems.
Of the 12 systems, we were not able to identify only one, and of the remaining 11 systems, only four are gravitationally bound.
One of the four, namely $\theta^{1}$~Ori, had already been discovered by Castelli, Galileo, and Huygens.
This fact does not diminish Hodierna's status as the first astronomer in history to publish a list of multiple systems, over a century earlier than previously accepted.

\section*{Acknowledgements}

We thank the reviewer Robert Argyle for his useful advices to improve the paper, Firas Sabia for extracting text from the original scanned document using artificial intelligence techniques, and Marco Longhitano for kindly responding to our inquiries.
We acknowledge financial support from the Agencia Estatal de Investigaci\'on of the Ministerio de Ciencia e Innovaci\'on and the ERDF ``A way of making Europe'' through project
PID2022-137241NB-C42 
and from the European Commission Framework Programme Horizon 2020 Research and Innovation through the ESCAPE project under grant agreement no.~824064.
This research made use of the NASA's Astrophysics Data System Bibliographic Services and the Exoplanet Archive, which is operated by the California Institute of Technology, under contract with the National Aeronautics and Space Administration under the Exoplanet Exploration Program, 
the Washington Double Star catalogue maintained at the U.S. Naval Observatory,
the Simbad database \citep{wenger00}, 
the VizieR \citep{ochsenbein00} catalogue access tool, and the Aladin sky atlas \citep{bonnarel00} at the Centre de donn\'ees astronomiques de Strasbourg (France).

\section*{Data Availability}

The data underlying this article are available in the article.



\bibliographystyle{mnras}
\bibliography{biblio}




\appendix

\section{Three binaries mentioned in the \textit{``Almagestum Novum''}}
\label{sec:riccioli}

Giovanni Battista Riccioli (1598--1671) was an Italian astronomer and Jesuit priest, famous for his experiments with pendulums and falling bodies, his analysis of 126 arguments about the Earth's motion, and for introducing the modern system of lunar nomenclature.
In his encyclopedic book \textit{``\textit{Almagestum novum astronomiam veterem novamque complectens observationibus aliorum et propriis novisque theorematibus, problematibus ac tabulis promotam in tres tomos distributam}''}\footnote{``Almagestum Novum, encompassing ancient and modern astronomy, advanced with observations of others and his own, as well as with new theorems, problems, and tables, distributed in three volumes''.}, \citet{riccioli1651} described three binaries hypothetically discovered by the Dutch astronomer and mathematician Martinus Hortensius (Martin van den Hove, 1605--1639).
In particular, when \citet{riccioli1651} portrayed the capacity of Hortensius' telescope to measure the diameter of stars, the former wrote:
``(...) and comparing this capacity with other distances, he found that the two contiguous stars in Capricorn are separated by one-eighth of the length of the tube, that is, 5 or 5\,{\textonehalf} arcminutes, and the two contiguous stars in the Hyades are nearly 5 or 4\,{\textonehalf} arcminutes'' (First volume, Sixth book, Chapter IX, Paragraph 4; translated from the original in Latin). 
We were not able to identify the original claim by Hortensius, nor a third double star in the Pleiades separated by astonishing 31\,arcmin.

Following the same procedure as described for Hodierna's double stars, we propose that the two first pairs of Hortensius and Riccioli were Algedi ($\alpha^{2}$~Cap) and Prima Giedi ($\alpha^{1}$~Cap) in Capricorn, which are separated by about 6.4\,arcmin in the sky and make up our system \#4, and $\theta^{2}$~Tau and $\theta^{1}$~Tau in the Hyades, which are separated by about 5.6\,arcmin.
Curiously, if these assignations are correct, Riccioli's estimations of angular separations were also affected by the same $\sim$ 0.80 scale factor as Hodierna's.


%
\section{Pairs thumbnails}
\label{sec:binary_star_images}

\begin{figure*}
  \centering
  \includegraphics[width=0.8\linewidth, angle=0]{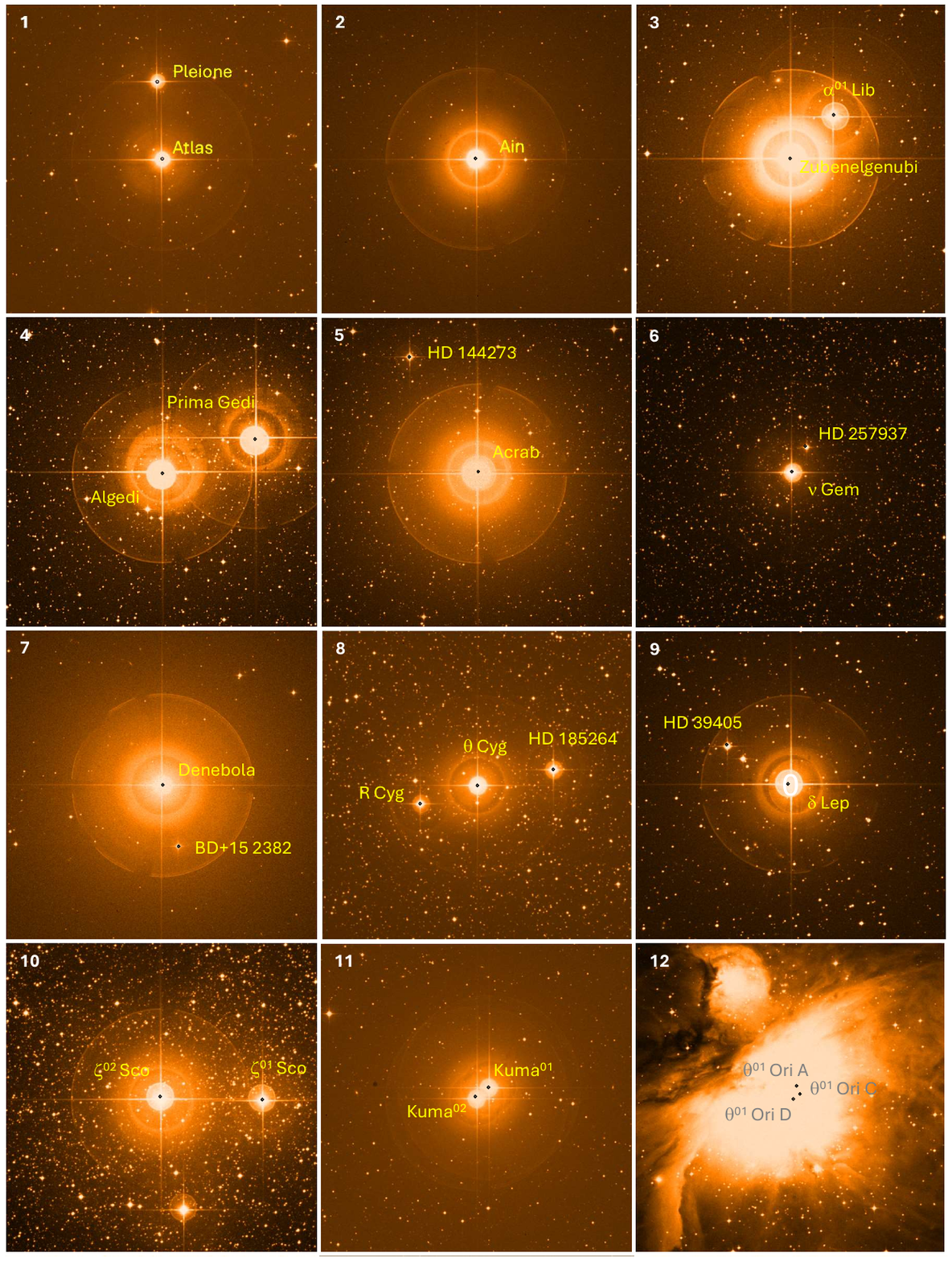}
 \caption[]{European Southern Observatory Digital Sky Survey images of Hodierna's double stars.
 All images are 20 $\times$ 20\,arcmin and DSS2-red except for image \#12, which is DSS2-infrared.}
  \label{fig:binary_stars_images}
\end{figure*}


\bsp	
\label{lastpage}
\end{document}